\documentstyle[preprint,tighten,floats,prd,aps]{revtex}
\input epsf
\newcommand{\be}{\begin{eqnarray}}
\newcommand{\ee}{\end{eqnarray}}

\begin{document}
\draft

\preprint{hep-ph/9712547 \hspace{111mm} UAHEP9716}

\title{Virtual Supersymmetric Corrections in $e^+e^-$ Annihilation}
\author{L. Clavelli\footnote{lclavell@ua1vm.ua.edu}, P.W. Coulter
\footnote{pcoulter@ua1vm.ua.edu}, and Levan R. Surguladze
\footnote{levan@gluino.ph.ua.edu}}

\address{Department of Physics and Astronomy, University of Alabama,
                         Tuscaloosa AL 35487}
\date{December 1997}
\maketitle

\begin{abstract}
Depending on their masses, Supersymmetric particles can affect various
measurements in Z decay.  Among these are the total width (or
consequent extracted value of $\alpha_s$), enhancement or suppression
of various flavors, and left-right and forward-backward asymmetries.
The latter depend on squark mass splittings and are, therefore, a
possible test of the Supergravity related predictions.  We calculate
leading order corrections for these quantities considering in particular
the case of light photino and gluino where the SUSY effects are
enhanced.  In this limit the effect on $\alpha_s$ is appreciable,
the effect on $R_b$ is small, and the effect on the
asymmetries is extremely small.
\end{abstract}


\renewcommand{\theequation}{\thesection.\arabic{equation}}

\renewcommand{\thesection}{\arabic{section}}

\section{\bf Introduction}
\setcounter{equation}{0}

     In recent years several slight anomalies in Z decay have
encouraged attention to the possibility of supersymmetric particles
in the Z region and below.  These are 1) an apparent value of the
strong coupling at the Z greater than expected from the extrapolation
of low energy measurements, 2) an enhanced value of the Z branching
ratio into b quarks relative to the Standard Model (SM) expectations,
and 3) deviations from the Standard Model predictions for the
left-right asymmetry in polarized electron-positron annihilation and for
various forward backward asymmetries.   Even if these discrepancies
shrink under further analysis as may now be the case with the $R_b$
anomaly, it is useful to consider what constraints they impose on
SUSY masses.  In this note we treat the strong supersymmetric (SUSY)
effects on these quantities from the graph of Fig.\ 1 involving a
virtual squark-squark-gluino triangle correction to the Z coupling to
quark-antiquark final states. We assume that, while the gluino might
be light relative to the Z mass, the squarks are all above the Z so
there are no effects from production of on-shell squarks.  The graph
of Fig.\ 1 has been considered earlier \cite{Altarelli,BR,Krasnikov,Djo}.
These works for the most part ignore sfermion mass splittings crucial
to some of the anomalies discussed above and, in any case, do not consider
explicit mass splitting schemes currently under discussion.
  The present work, extending the results of these authors
to include these effects, is divided as follows.  In section 2 we
briefly discuss the experimental situation relative to the
above-mentioned anomalies.  In section 3 we present the calculation
of the graph in Fig.\ 1, allowing for squark mass splittings.
 Results and conclusions are given in section 4.

\section{\bf Anomalies in Z Decay}
\setcounter{equation}{0}

     Since 1992, various authors have noted that the apparent value of
the strong coupling constant at the Z is greater than the value
expected in the Standard Model from extrapolation from lower energies.
This was in fact one of the initial proposed positive indications
\cite{LC,Hebbeker,Jezabek} of a light gluino since such a particle
would slow the running of $\alpha_s$.  The effect has been clouded by
disagreement over the value of this coupling at both the low and high
energy regions but, clearly, the greater is the value of
$\alpha_s(M_Z)$ and the lower is its value at lower
energies, the more likely one is to be interested in the light gluino
possibility.  The fitted values quoted from Z decay measurements
have ranged from a low of $0.108 \pm .005$ \cite{Delphi} to a high of
$0.141 \pm .017$ \cite{LEP92}.  Current values are typified
\cite{PDG96} by that from the Z hadronic width.
\be
      \alpha_s(M_Z)^{\mbox{\scriptsize apparent}} = 0.123 \pm 0.005
\label{alphasexp}
\ee
Here we have labeled the fitted value as "apparent" since the
value quoted is from a Standard Model analysis and there could be
extra (e.g., SUSY) contributions such as that of Fig.\ 1.  We assume
that the actual value of $\alpha_s(M_Z)$ is that extrapolated from
low energy.  Although these numbers are somewhat dependent on the
analysis performed, they tend to be systematically below the value in
Eq.~(\ref{alphasexp}).  For definiteness we will note the values from
\cite{PDG96} in the SM or heavy gluino (HG) case as obtained from
an $\Upsilon$ decay analysis

\begin{figure}
\hskip 3.0cm
\epsfxsize=3.5in \epsfysize=2in \epsfbox{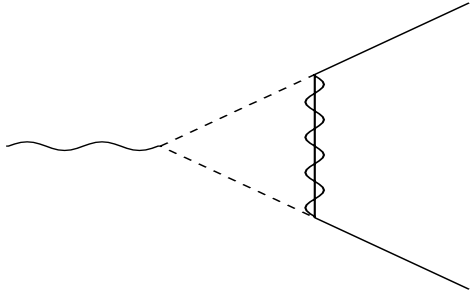}
\caption{Leading order virtual squark-squark-gluino correction to hadronic Z width.
Solid and dashed lines denote quarks and squarks correspondingly. Wavy line denotes
gluino}
\label{Fig.1}
\end{figure}
\be
   \alpha_s(M_Z)^{\mbox{\scriptsize SM}} = 0.108 \pm .010
\ee
and from \cite{CC} in the light gluino (LG) case
\be
   \alpha_s(M_Z)^{\mbox{\scriptsize LG}} = 0.114 \pm .005.
\ee
Thus the difference between the actual $\alpha_s(M_Z)$ and the
apparent value is
\be
         \delta \alpha_s = - 0.009 \pm .007  \mbox{ (LG case) }\\
         \delta \alpha_s = - 0.015 \pm .011  \mbox{ (SM or HG case) }
\ee
The SM value of the coupling constant adopted here is in agreement
with that found from QCD sum rules \cite{Shifman}, although these
authors quote large errors.
Other particular
analyses, notably that from $\tau$ decay, could significantly reduce
the discrepancy (or even change it's sign) but we feel it fair to
say that a large body of low energy data prefers negative $\delta
\alpha_s$.  The $\tau$ analysis in particular has been criticized
\cite{AltNasRid} as being especially vulnerable to non-perturbative
effects. In the LG case there is a small positive
contribution to $\delta \alpha_s$ from gluino contributions alone
to Z decay \cite{CCS} amounting to $\delta \alpha_s \approx +.002$.
Related perhaps to the $\alpha_s$ anomaly, is a possible excess
of b quark production in Z decay which has stimulated significant
interest in possible manifestations of SUSY below the Z.  In
particular it has been noted that an appreciable enhancement of b
production could be a signal for charginos and stop quarks below the Z
\cite{BR,Kane}.  Such a non-standard-model contribution to Z decay
might also explain the apparent enhancement in $\alpha_s$ at the Z
relative to the value expected from extrapolation from lower energies.
If the only non-SM contribution to the Z width was in the 
$b \overline{b}$ channel it would be related to $\delta \alpha_s$ by
 \cite{LC95}
\be
       \delta \alpha_s = \pi \delta R_b { {1+ \alpha_s/\pi} \over
                  {1 - R_b} },
\ee
where
\be
      R_b = { {\Gamma (Z \rightarrow b \overline{b})} \over
               {\Gamma (Z \rightarrow \mbox{hadrons})} }
\ee
and $\delta R_b$ is the deviation of this quantity from its standard
model value.  However, due to new LEP results above the Z,
the likelihood of low mass charginos (at least in the standard SUSY
picture) has diminished.  In addition the b anomaly has largely
disappeared under reanalysis.  The latest experimental analysis 
yields \cite{ALEPH97}
\be
   \delta R_b = .0009 \pm .0011
\ee
This value is consistent with zero and it is interesting to see how
severely it constrains contributions beyond the Standard Model and, in
particular, whether SUSY effects could lead to an enhanced apparent
$\alpha_s$ without enhancing the b decay mode of the Z.
     With sufficiently light chargino and stop quarks the 
chargino-stop virtual correction to the Z vertex dominates the SUSY
contribution to the Z partial widths and can lead to a $\delta R_b$ as
great as .0018.  In the light gaugino scenario, however, with
$m_{1/2}=A=0$, the stop quarks are expected to be above the top
\cite{CG} and this contribution to $\delta R_b$ drops below .0005.  In
this case, the virtual gluino correction could become the dominant
SUSY effect.  In this article we reconsider the gluino (and photino)
corrections with special attention to scheme dependence and dependence
on the sfermion mass splitting which have not been part of prior
investigations.

\par
     We also consider the left-right and
forward-backward asymmetries.  These are related to the effective
coupling of the Z to vector and axial vector currents of particular
fermion species.  One defines
\be
    A(f) = {{2 g_V(f) g_A(f)} \over{g_{V}^2(f)+g_{A}^2(f)}}.
\ee
The left-right asymmetry is defined in terms of the total annihilation
cross sections for left-handed and right-handed polarized electrons.
\be
 A_{LR} = {{\sigma_L - \sigma_R} \over {\sigma_L + \sigma_R}} = A(e).
\ee
The Forward Backward asymmetry for a particular flavor f is
\be
 A_{FB}(f) = {{\sigma(\cos\theta >0)
 - \sigma(\cos\theta <0)} \over {\sigma(\cos\theta >0)
      + \sigma(\cos\theta <0)}}.
\ee
For an unpolarized beam this is given by
\be
     A_{FB}(f) = {3 \over 4} A(e) A(f).
\label{AFB}      
\ee
     Current asymmetry measurements can be summarized by experimental
determinations of the difference between the $A(f)$ and their standard
model predictions \cite{PDG96}.
\begin{displaymath}
   \delta A(e) = 0.011 \pm .005
\end{displaymath}
\be
\hspace{14mm}   \delta A(b) = -.093 \pm .053
\ee
\begin{displaymath}
   \delta A(c) = -.061 \pm .090
\end{displaymath}
Even if, as we find, the SUSY contributions cannot account for
discrepancies at this level, it is useful to record here the 
SUSY prediction for comparison with future more accurate measurements.  

\section{\bf SUSY QCD Vertex Correction}
\setcounter{equation}{0}

     The Born term matrix element for Z decay into a
fermion-antifermion pair in the Standard Model takes the form

\be
     M^0_{Z\rightarrow f\overline{f}} = {e \over{ \sin 2 \theta }} \epsilon_\mu
          \overline{u}(p) \gamma_\mu (g_V^0-g_A^0\gamma_5)v(q-p),
\ee
where
\be
      g_V^0 = T_3 - 2 Q \sin^2\theta \quad , \quad g_A^0 = T_3.
\ee
In terms of these parameters the zeroth order Z width into a fermion
anti-fermion pair, neglecting the fermion mass, is
\be
 \Gamma^0_{f\overline{f}} = { \alpha \over \sin^2 2\theta  }
                               M_Z [(g_V^0)^2+(g_A^0)^2].
\ee
A second quantity of some interest is
\be
     A^0(f) = { 2g_V^0 g_A^0 \over (g_V^0)^2+(g_A^0)^2 }
\ee
since the forward-backward asymmetry of fermion species $f$ is given by
eq.\ (\ref{AFB}), 
and the left-right asymmetry of the total hadronic cross section with
a longitudinally polarized electron beam is $A_{LR} = A^0(e)$.
The existing discrepancies between these measured quantities and the
Standard Model predictions make a study of the supersymmetric
contributions of interest.  The lowest order SUSY QCD corrections to
the Z-fermion vertex arise from the squark-squark-gluino triangle
graph of Fig.\ 1.  The corresponding matrix element is
\be
 \delta M_{Z\rightarrow f\overline{f}} = {e \over{ \sin 2 \theta }}
         \epsilon_\mu
          \overline{u}(p) 
           \gamma_\mu (\delta g_V^0-\delta g_A^0\gamma_5)v(q-p),
\ee
where
\be
  \delta g_V = [ g_V^0(I_L^r+I_R^r) + g_A^0(I_L^r-I_R^r) ]
                   C_F {\alpha_s(\mu) \over{4 \pi}},
\ee
\be
  \delta g_A = [ g_A^0(I_L^r+I_R^r) + g_V^0(I_L^r-I_R^r) ]
                   C_F {\alpha_s(\mu) \over{4 \pi}},
\ee
and $C_F$ is the eigenvalue of the quadratic Casimir operator of the
gauge group in the fundamental representation (for SU(3), $C_F=4/3$).
The $I^r_{L,R}$
are the contributions from left and right squarks corresponding to the
graph of Fig.\ 1.  The $r$ indicates that the ultraviolet renormalization
has been made. Adding the SUSY correction to the Born term matrix element is
equivalent to replacing $g^0_{V,A}$ by
\be
      g_{V,A}=g^0_{V,A} + \delta g_{V,A}.
\ee
These lead to the $O(\alpha_s)$ equations
\be
    g^2_V(f)+g^2_A(f) = [ (g^0_V(f))^2+(g^0_A(f))^2 ] \biggl\{
          1 + C_F {\alpha_s(\mu) \over{2 \pi}} \biggl[ (I_L^r+I_R^r)
                + A^0(f)(I_L^r-I_R^r)\biggr] \biggr\}
\ee
\be
    A(f) = A^0(f) + C_F {\alpha_s(\mu) \over{2 \pi}}
                                 \biggl(I_L^r-I_R^r \biggr)
\ee
Using dimensional regularization \cite{tHooft} with the dimension of
space - time d$=4-2\epsilon$ and a standard Feynman parametrization,
the unrenormalized $I_{L,R}$ can be written as follows.
\begin{displaymath}
 I_{L,R} = { 16i  (2 \pi)^{2-d} \mu^{2 \epsilon} \over d-4 }
         \int d^dk \int_0^1 dx_1 dx_2 \theta(1-x_1-x_2)
    { C_{L,R} \over (C_{L,R}-k^2)^3 } 
\end{displaymath}
or
\be
 I_{L,R} = \Gamma(\epsilon) (4 \pi)^\epsilon \int_0^1
 ({C_{L,R} \over {\mu^2}})^{-\epsilon} dx_1 dx_2 \theta(1-x_1-x_2),
\ee
where
\be
    C_{L,R} = {\tilde \mu}^2 x_2 + {\tilde m}^2_{L,R} (1-x_2)
                              - Q^2x_1(1-x_1-x_2).
\ee
Here $\tilde{\mu}$ is the gluino mass and $\tilde{m}_{L,R}$
are the left and right squark masses of the same flavor as the final
state fermions.  $Q^2$ is the center-of-mass energy squared
($M^2_Z$ at the Z pole) and $\mu$ is t'Hooft's unit of mass
\cite{tHooft}.

Adding SUSY self energy corrections on the external quark legs has the effect of subtracting
from the result of Fig.1 its value at $Q^2=0$ \cite{Djo}. Then
\be
  I_{L,R} = - \int _0^1 dx_1 dx_2
       \theta(1-x_1-x_2) \ln \biggl[1 - {Q^2 x_1  (1-x_1-x_2)
                 \over
          \tilde{\mu}^2 x_2 + \tilde{m}^2_{L,R}(1-x_2) }\biggr].
\ee
This function has the series expression
\be
     I^{CWZ}_{L,R} = \sum_{n=1}^\infty
     {{Q^{2n}} \over n} B(n+1,n+1) \int
               _0^1 dx_2 {{(1-x_2)^{2n+1}}\over{ [\tilde \mu}^2 x_2
                + {\tilde m}^2_{L,R} (1-x_2)]^n}.
\ee
Two simple cases can be studied:
\be
   I_{L,R}({\tilde \mu}=0) = \sum_{n=1}^\infty
   \biggl({ Q^2 \over \tilde{m}^2_{L,R} }\biggr)
  { B(n+1,n+1) \over n(n+2) }
    = {1 \over 18}\biggl({ Q^2 \over \tilde{m}^2_{L,R} }\biggr)
      + {1 \over 240 }\biggl( { Q^2 \over \tilde{m}^2_{L,R} }\biggr)^2
                   + \cdots
\ee
and
\be
  I_{L,R}(\tilde{\mu}=\tilde{m}_{L,R})=\sum_{n=1}^\infty
    \biggl({ Q^2 \over \tilde{m}^2_{L,R} }\biggr)^n
  { B(n+1,n+1) \over n(2n+2) } 
     = {1 \over 24 } { Q^2 \over \tilde{m}^2_{L,R} } + \cdots
\ee
In each case the series are rapidly converging and the first term gives
about $85\%$ of the full result even down to $\tilde{m}^2_{L,R} = Q^2$.
One can see from these results that although $\hat{I}$ falls rapidly
with increasing squark mass it is much less sensitive to the gluino
mass.  Since an ultra-light gluino is still a possibility
\cite{Farrar}, in the following section we concentrate on the
numerical results for the $\tilde{\mu}=0$ case.
     The one-loop SUSY QCD corrected decay rates and asymmetries
into $f\overline{f}$ are to  $O(\alpha_s)$
\be
   \Gamma_{f\overline{f}} = { \alpha \over \sin^2 2 \theta }
         M_Z (g_V^2+g_A^2)
        = \Gamma^0_{f\overline{f}} \biggl[ 1 
           +{3 \over 4} C_F { \alpha_s \over \pi } +
           C_F { \alpha_s \over \pi }
                \biggl( { I^r_L+I^r_R \over 2 }
            + A^0(f) { I^r_L-I^r_R \over 2 }\biggr)
\label{eq:Gff}
\ee
\be
   A(f) = { 2 g_V g_A \over g_V^2+g_A^2 } = A^0(f) 
       + C_F {\alpha_s \over \pi } { I^r_L-I^r_R \over 2 }
\label{eq:Af}
\ee
In $\Gamma_{f\overline{f}}$ we have included the QCD order $\alpha_s$ correction.
The total hadronic decay rate to this order, obtained by summing the
partial rates over $f$, can be written as follows.
\be
    \Gamma_{\mbox{\scriptsize tot}} = \Gamma^0_{\mbox{\scriptsize tot}}
     \biggl(1 
     + { \alpha_s^{\mbox{\scriptsize apparent}} \over \pi }\biggr),
\ee
where the "apparent" $\alpha_s$, the one that would
result from a Standard Model analysis, is
\be
\delta \alpha_s = \alpha_s - \alpha_s^{\mbox{\scriptsize apparent}}
         =\pi \sum_f \delta_f
\ee
where
\be
  \delta_f = C_F { \alpha_s \over 2 \pi }
     { \Gamma_{f\overline{f}} \over \Gamma_{\mbox{\scriptsize tot}} }
       [I^r_L(f)+I^r_R(f) + A^0(f)(I^r_L(f)-I^r_R(f))].
\ee
$\Gamma^0_{\mbox{\scriptsize tot}}$ being the zeroth order total hadronic decay width.

The contribution to $R_b$ is
\be
  \delta R_b = \delta_b - R^0_b \sum_f \delta_f.
\ee
with the analogous expression for $\delta R_c$.
The $I_{L,R}$ depend on the masses of the SUSY partners of the
left and right handed quarks of flavor f as well as on the gluino
mass.  The squark masses, in the SUGRA scheme are given by

\be
    {\tilde m}^2_{f,L} = m^2_0 + m^2_f + M^2_Z \cos 2 \beta
        \bigl ( T_{3,f} - Q_{f} \sin^2 \theta \bigr )\\
    {\tilde m}^2_{f,R} = m^2_0 + m^2_f + M^2_Z Q_{f} \cos 2 \beta
                                                          \sin^2 \theta
\ee

where $m_0$ is the universal scalar mass.  Neglecting the quark
masses, and $M_Z^2$, the $I_{L,R}$ become independent of flavor.  In
this case the lowest order contributions to $\delta R_b$ and to the
asymmetries vanish although there is still a significant effect on the
Z total width.

\section{\bf Results and Conclusions}
\setcounter{equation}{0}

Since the experimental evidence for discrepancies from the Standard Model
are so slight (two standard deviations or less), at this point
our results should be interpreted as predictions of low energy SUSY
rather than as suggestions from the data. Our main point is that the
graph of Fig.\ 1 can be effective in enhancing the hadronic decay rate 
of the Z without enhancing the relative contribution from b quarks.
The variations of $\delta R_b$ and $\delta \alpha_s$ are shown
 in Figs.\ 2 and 3 respectively as functions
of the universal scalar mass $m_0$, using $\alpha_s(M_Z)=0.113$,
$\tan \beta=1.6$ and $\sin^2 \theta = 0.2317$.
The contribution to $\delta R_b$ and to
$\delta \alpha_s$ are negative and slowly varying.  The SUSY QCD
contribution to $\delta R_b$ therefore tends to cancel the
chargino-stop contribution to $\delta R_b$ while still leaving an
appreciable negative contribution to $\delta \alpha_s$ consistent with
the experimental indications discussed in section 2.   The
contributions to the forward-backward asymmetries behave similarly
although they are orders of magnitude smaller than the current
experimental errors.  The SUSY predictions are very close to
those of the Standard Model and are two standard deviations from the
current experimental results.
\begin{table}
\caption{Results for different schemes.}
\smallskip
\begin{tabular}{|l|l|l|l|l|l|l|l|l|}
 $f$ & $\tilde{m}_L$ & $\tilde{m}_R$ & $\Gamma^0$ &
               $\delta\Gamma \times 10^4$ &
        $\delta R_f \times 10^5 $ &  $A^0$ & $\delta A \times 10^4$ &
                                             $\delta A_{\mbox{\scriptsize exp}}$  \\
                                                                         \hline
    &     &     &      &       &      &      &     &             \\
  $c$ & 100 & 103 & .287 & 6.73 & + 5.78 & .667 & 0.826 & .06$\pm$.09 \\
     &     &     &      &       &       &      &      &              \\
  $b$ & 113 & 107 & .370 & 6.78 & -3.85 & .935 &-1.06 & .10$\pm$.05   \\ \hline  
\end{tabular}
\label{results}
\end{table}
  In table 1 we summarize the results in the case $m_0=106 GeV$ and
$\tilde \mu \approx 0$ as suggested in recent phenomenological studies of the
Fermilab jet transverse energy distributions, scaling ratio, and top
quark events \cite{CT,CG}. $\delta A_{\mbox{\scriptsize exp}}$ is defined as
$A_{\mbox{\scriptsize exp}}-A^0$ and $\delta \Gamma$ is the SUSY correction to
the Z width from eq.\ (\ref{eq:Gff}). The left-right asymmetry is
obtained by  replacing $C_F \alpha_s$ by $\alpha$ in eq.\ (\ref{eq:Af}).
Since the Standard Model
result is suppressed by a factor of $4 \sin^2 \theta - 1$ one could expect
the experiment to be sensitive to the SUSY contribution.  However in
the SUGRA related model for sfermion mass splittings, the selectron
mass splitting is also suppressed by the same factor.  This fact plus
the proportionality to the fine structure constant makes this
experiment very insensitive to SUSY.

     The analysis presented here can be readily extended to the higher
$Q^2$ values of LEP II.  The early results from CERN are consistent
with the Standard Model but, in view of the large fluctuations seen in
the early results from LEP I discussed in the introduction to this
article and in view of the sharply reduced statistics at higher energy
and complications from the "radiative return" and W pair production we
prefer to defer this analysis to a later date.   Parity violating
effects in hadronic collision from SUSY virtual contributions such as
those treated here in Z decay have also received recent attention
\cite{CPYUAN}.

\begin{figure}
\hskip 1.0cm
\epsfxsize=7in \epsfysize=6in \epsfbox{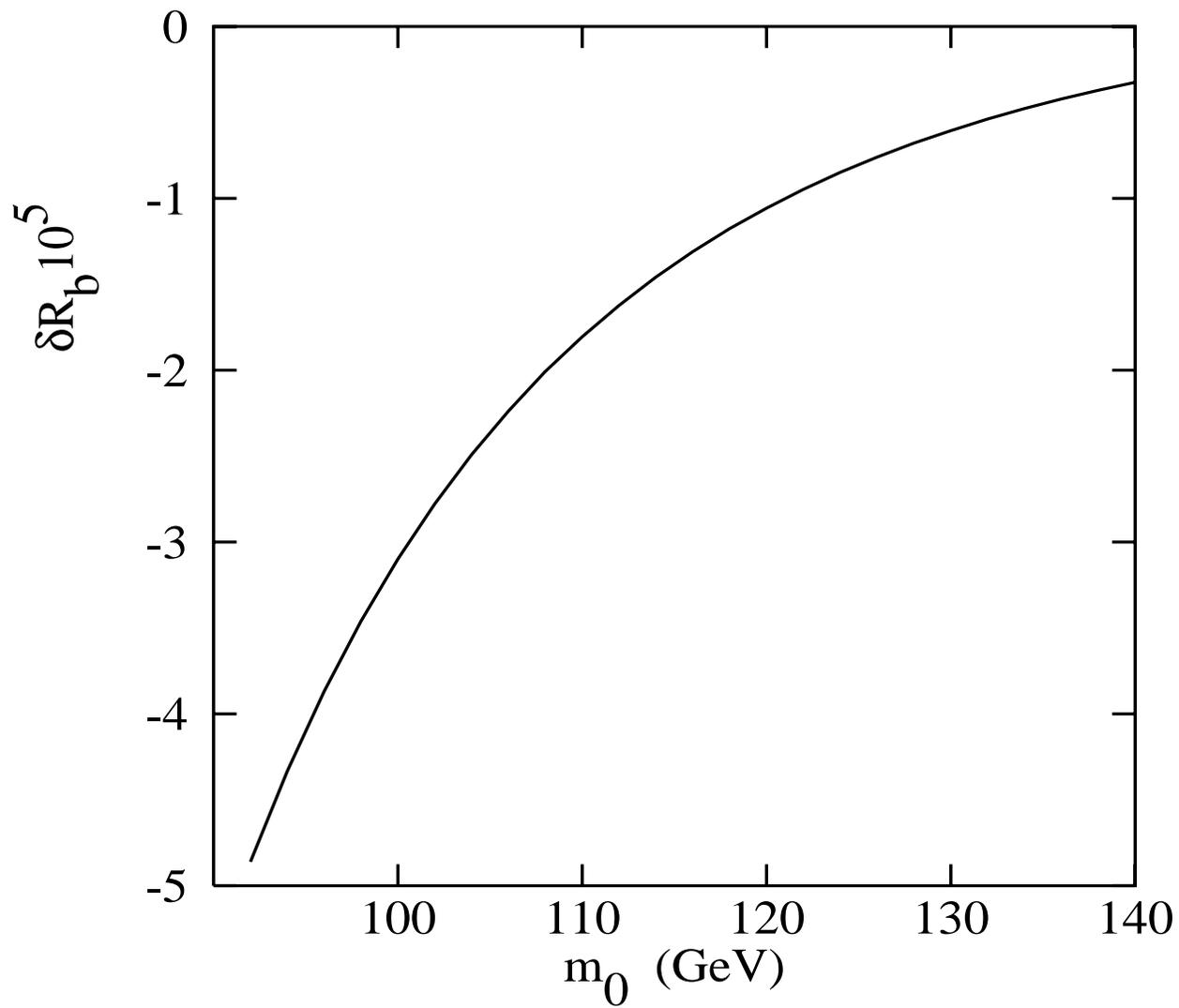}
\caption{Results for different schemes vs universal scalar mass}
\label{Fig.2}
\end{figure}
\begin{figure}
\hskip 1.0cm
\epsfxsize=6in \epsfysize=5in \epsfbox{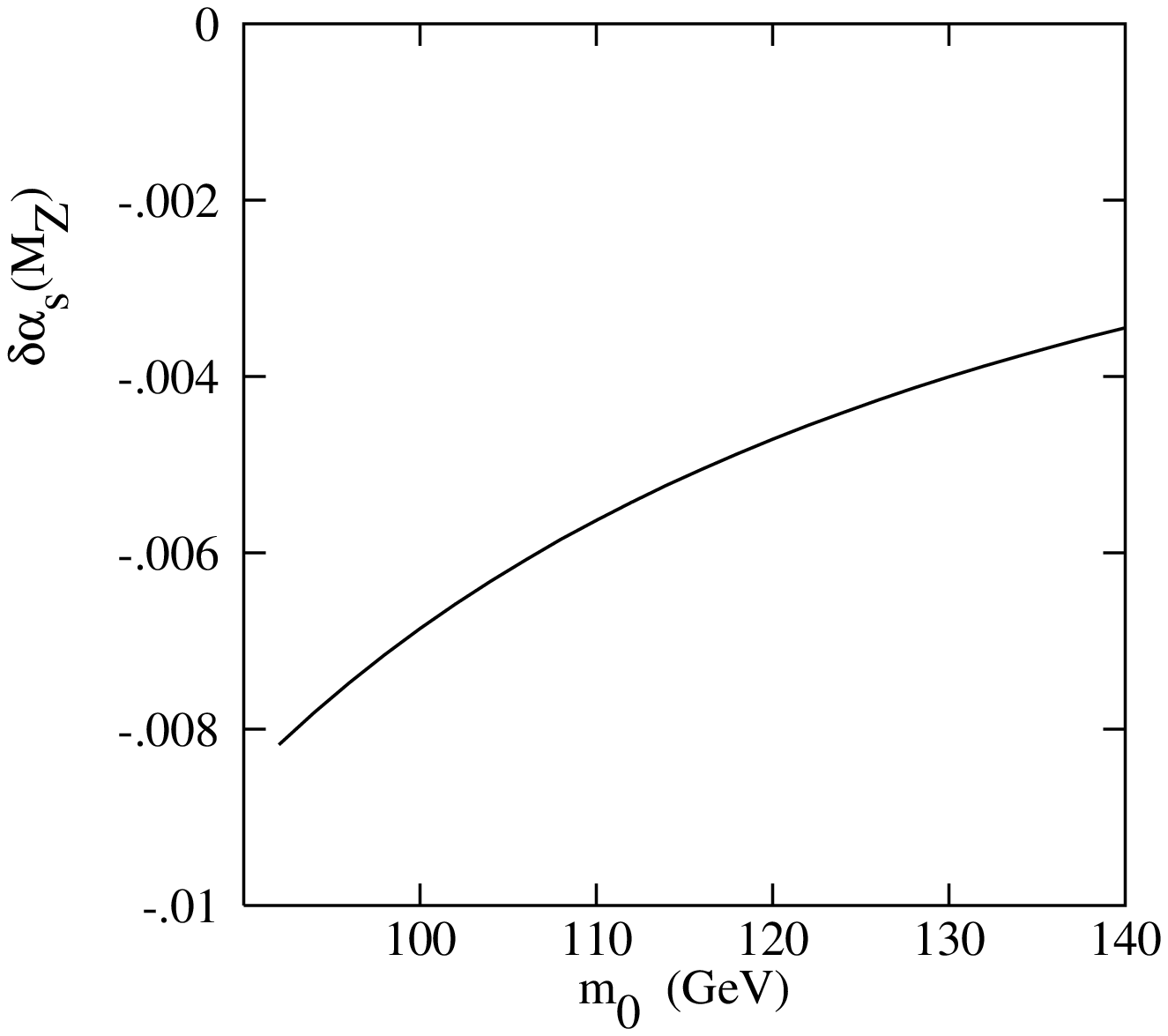}
\caption{Effect in $\alpha_s$ vs universal scalar mass}
\label{Fig.3}
\end{figure}

\acknowledgements

The authors would like to thank Peter Povinec for participating
in the early stages of this work and A.Djouadi for communication
on his results \cite{Djo}.
This work was partially supported by the U.S. Department of Energy
under grant No. DE-FG05-84ER40141.

\end{document}